\documentclass[pra,twocolumn,superscriptaddress]{revtex4}
\usepackage{blindtext}
\usepackage{amsfonts}
\usepackage{amsmath,amssymb}
\usepackage{graphicx}
\usepackage{subfigure}
\usepackage{float}
\usepackage{bm}
\usepackage{epstopdf}
\usepackage{url}
\usepackage{hyperref}
\hypersetup{
     colorlinks = true,
     linkcolor = blue,
     anchorcolor = blue,
     citecolor = blue,
     filecolor = blue,
     urlcolor = blue
}

\begin{document}

\title{Magnetic domains in 2D moir\'e lattices with square and hexagonal symmetry}
\author{C. Madro\~nero and R. Paredes} 
\affiliation{Instituto de F\'{\i}sica, Universidad
Nacional Aut\'onoma de M\'exico, Apartado Postal 20-364, M\'exico D.F. 01000, Mexico.}  
\email{rosario@fisica.unam.mx}

\begin{abstract}
We report the persistence of magnetic domains lying in moir\'e patterns with square and hexagonal symmetries.  Our investigation is based on the dynamical description of two magnetic domains represented by a two species bosonic mixture of $^{87}$Rb ultracold atoms, being each specie initially localized in the left and right halves of a moir\'e lattice defined by a specific angle $\theta$. To demonstrate the persistence of such initial domains, we follow the time evolution of the superfluid spin texture, and in particular, the magnetization on each halve. The two-component superfluid, confined in the moir\'e pattern plus a harmonic trap, was described through the time dependent Gross-Pitaevskii coupled equations for moir\'e lattices having $90 \times 90$ sites. Results showed the existence of rotation-angle-dependent structures for which the initial magnetic domain is preserved for both, square and hexagonal moir\'e patterns; above $\theta >10^\circ$ the initial magnetic domain is never destroyed. Stationary magnetic states for a single component Bose condensate allowed us to identify the lattice parameter associated with moir\'e crystals that emerge for twisting angles belonging to the intervals $\theta \in \left(0^\circ ,30^\circ \right)$ and $\theta \in \left(0^\circ ,45^\circ \right)$ for hexagonal and square geometries respectively.
\end{abstract}

\maketitle

\section{Introduction}

The investigation here addressed is connected with the phenomenology and physical properties arising in van der Waals structures. In particular, with the  magnetic properties persisting in patterns formed when two periodic lattices lie on top of each other with a relative twist between them. That is, moir\'e structures supporting magnetic domains for long times. In recent years exotic transport and magnetic properties have been measured in assemblies of monolayers of transition-metal dichalcogenides \cite{WangY,Ponomarenko,Dean,Hunt}. It is believed that as a result of the incommensurate lattice structure arising from the interlayer twist, and/or lattice constant mismatch, anomalous interlayer couplings are originated, and consequently,  profound effects on the transport and optical properties in the bilayer arrays arise. This novel class of 2D materials exhibit ultrafast interlayer charge transfer that facilitates the photocurrent generation and the formation of interlayer excitons. Even more, twisted bilayer graphene has been demonstrated to exhibit superconductivity for the so called magic angles of rotation \cite{Cao}, correlated insulating behavior \cite{Cao2} and magnetization textures in 2D magnets \cite{Xiao}. As previously reported in the literature, the understanding of the strength and the form of the interlayer coupling in the van der Waals structures is the essential element that gives light of the interesting phenomena observed in homo- and hetero-bilayer lattices.

The theoretical study of properties rising in twisted bilayers has been addressed from several routes. While the emergence of superconductivity for the magic angles was explained in terms of a continuum model that predicts flats bands \cite{Tarnopolsky}, the emergence of ferromagnetism was predicted in twisted bilayer graphene using first-principles density functional theory calculations \cite{Yndurain, Lopez-Bezanilla}. An effective spin model advises that a system is described by a ferromagnetic Mott insulator whenever maximally localized superlattice Wannier wave functions are present \cite{Kangjun}. The phase diagram at mean field level for a moir\'e-Hubbard model has revealed a variety of phases including Wigner crystals with charge-density wave forms and Chern insulators \cite{Pan}. Perhaps the most common feature found in those theoretical investigations, is the existence of a particular angle of rotation among graphene layers for which phases and particular properties are displayed. 

In the present study we concentrate in tracking the dynamics of a double magnetic domain that evolves under the influence of a moir\'e pattern. The double magnetic domain is represented by two different hyperfine spin components of a Bose-Einstein condensate, placed in the two dimensional optical pattern \cite{GTudela}. The lattice where this double-magnetic domain lies, can be generated by superimposing two rotated optical lattices with a definite symmetry. The particular moir\'e patterns that we consider emerge from the superposition of two square- and two hexagonal- lattices in 2D. As recently reported in the literature, experiments with ultracold clouds of $^{87}$Rb neutral atoms confined in 2D optical lattices are an ideal candidate where twistronics can be envisaged. In such a system Bose-Einstein condensates loaded into spin-dependent optical lattices form the moir\'e structure that might serve to investigate the physics underlying the superconductivity in twisted-bilayer-graphene \cite{Zengming}. Other techniques through which moir\'e patterns can be created are by means of laser interference lithography \cite{Mahood}, or using digital micromirror devices \cite{Choi, Takahashi}.

The main interest of this investigation is to establish how magnetic properties can be controlled by changing the twisting angle between two lattices with either, square or honeycomb symmetries. Our particular interest is to establish the angles for which a given initial state composed of a double magnetic domain prevails. To accomplish such a purpose we prepared as the initial condition a non stationary state, and follow its dynamical evolution for sufficiently long times. This initial state is formed by two different hyperfine spin components of a Bose-Einstein condensate, such that initially each specie lies on left and right halves of a moir\'e pattern defined by a specific twist angle. Because of miscibility of the components was guaranteed by properly choosing values of the intra and inter-species interaction, the domains were expected to vanish. The shortest and largest size of the lattices for which our numerical experiments were performed are $40 \times 40$ and $90 \times 90$ sites. On one side, the stationary analysis for a single spin component allowed us to identify the variety of superlattices that emerge as a function of the rotation angle, as well a recognizing the quasicrystalline patterns appearing for definite twist angles. And, on the other side, we reached the conclusion that the initial ferromagnetic state prevails for angels above $\theta >10^\circ$ disregarding the hexagonal or square geometry respectively.  

The paper is structured as follows. First in Sec. \ref{Model} we present  the coupled equations that characterize the dynamics of two hyperfine spin components. Then, in Sec. \ref{Initialstates} we explain how to generate the initial state composed of two magnetic domains lying in definite moir\'e patterns. Afterwards, in Sec.
\ref{crystals} we present the analysis associated with the crystalline moir\'e structures that emerge from the rotated bilayers. In Sec. \ref{dynamics} we discuss the dynamics developed by the ferromagnetic state as a function of the angle that defines a given moir\'e pattern. Finally, we summarize our findings in Sec. \ref{Conclusion}, and provide an outlook of future directions of our paper.

\section{Two-component weakly interacting Bose gas confined in Moir\'e lattices}
\label{Model}

The aim of this investigation is to describe the dynamics of two magnetic domains evolving under the influence of weak interactions among its constituents, and lying in moir\'e lattices. The moir\'e structures that we shall consider, result from superimposing a pair of 2D rotated square or hexagonal lattices. Our particular interest is to establish the persistence of the simplest bi-component magnetic domain lying in a moir\'e structure. The initial magnetic domains that we consider will be set, as described in subsection \ref{Initialstates}, with two halves in which a couple of different spin components are placed. For this purpose we shall consider an analogous system that, as described in the introduction, belongs to ultracold matter context; a weakly interacting $F=1$ spinor Bose condensate confined in a 2D optical lattice having a moir\'e like structure. Particularly, we concentrate in the dynamics of two hyperfine spin components, $|\uparrow \rangle= |F=1,m_F=-1\rangle$ and $|\downarrow \rangle=|F=2,m_F=-2\rangle$, lying in the 2D moir\'e lattices, represented by $V_{ \mathrm{ext}}\left(\vec {r}\right)$. Within the mean-field formalism the wave functions $\Psi_{\uparrow}$ and $\Psi_{\downarrow}$of the two species $|\uparrow \rangle$ and $|\downarrow \rangle$ respectively, obey the following effective coupled GP equations: 

\small
\begin{eqnarray} 
i\hbar \frac { \partial \Psi _{\uparrow} (\vec {r},t)}{ \partial t } =\left[ H_0(\vec {r}) +  g_{\uparrow\uparrow}|\Psi_{\uparrow}|^{2} + g_{\uparrow\downarrow}|\Psi_{\downarrow}|^{2} \right]  \Psi_{\uparrow}(\vec{r} ,t)\cr
i\hbar \frac { \partial \Psi _{\downarrow} (\vec {r},t)}{ \partial t } =\left[ H_0(\vec {r})  +   g_{\downarrow\downarrow}|\Psi_{\downarrow}|^{2} + g_{\downarrow\uparrow}|\Psi_{\uparrow}|^{2} \right]  \Psi_{\downarrow}(\vec{r} ,t),
\label{coupledGP}
\end{eqnarray}
where $H_0(\vec {r})= -\frac { \hbar^ 2 }{2m} \nabla_{\perp}^ 2 +V_{ \mathrm{ext}}\left(\vec {r}\right)$ with $\nabla_\perp^2=\frac{\partial^2}{\partial x^2}+\frac{\partial^2}{\partial y^2}$ is the Laplacian operator in $2$D, $m$ the equal mass of the two spin components, and $\vec {r}= x \hat i +y \hat j$. The external potential in 2D has the following form:

\begin{equation}
\begin{split}
V_{ \mathrm{ext}}(x,y)=\begin{cases}
V_{HO}(x,y)+\frac{1}{2}\left[V_{sq}(x,y)+V_{sq}(x',y')\right]\\
V_{HO}(x,y)+\frac{1}{2}\left[V_{hx}(x,y)+V_{hx}(x',y')\right]
\end{cases}
\end{split}
\label{Ext_pot}
\end{equation}
where $V_{HO}(x,y)=\frac{1}{2}m\left ( \omega _x ^2 x^2 + \omega _y ^2 y^2\right )$ is a harmonic oscillator, that is usually present when an ultracold gas is produced in a laboratory. The values of the frequencies that we shall consider for our analysis are $\omega_x=\omega_y= \omega_r$, being $\omega_r/\omega_0=0, 0.4, 0.7$ and 1, with $\omega_0=2 \pi \times 50$ rad/s, this value of the frequency is a typical one in experiments with ultracold atoms \cite{Hadzibabic, Hung}. The variables $x'$ and $y'$ belong to a rotated system an angle $\theta$ as follows,
\begin{equation}
\left(\begin{array}{c}
x'\\
y'
\end{array}\right)=\left(\begin{array}{cc}
\cos\theta & -\sin\theta\\
\sin\theta & \cos\theta
\end{array}\right)\left(\begin{array}{c}
x\\
y
\end{array}\right).
\end{equation}
The second contribution of Eq. \ref{Ext_pot} is the term from which emerges the square and hexagonal moir\'e lattices, being each lattice given by,  
\begin{equation}
\begin{split}
&V_{sq}(x,y)={V}_{0}\left[{\cos}^{2}(kx)+{\cos}^{2}(ky)\right],\\
V_{hx}(x,y)={V}_{0} & \left[\cos(\frac{4ky}{3})+\cos(\frac{2kx}{\sqrt{3}}-\frac{2ky}{3})+\cos(\frac{2kx}{\sqrt{3}}+\frac{2ky}{3})\right],
\end{split}
\label{redes}
\end{equation} 

Regarding the values of the effective interaction couplings $g_{\sigma\sigma'}$ with $\sigma, \sigma' = \{ \uparrow, \downarrow \}$, they are written in terms of the $s$-wave scattering length $a_{\sigma, \sigma'}$ as, $g_{\sigma\sigma'}= 4\pi N\hbar^{2}a_{\sigma\sigma'}/m$, being $N$ the number of particles in the condensate. Its is important to stress here that since originally GP equation describes the ground state of the condensate in 3D, these interaction coefficients must be substituted by effective interaction couplings that take into account that the atom collision processes occur in 2D \cite{U2D, Salasnich, Mateo, Bao, Trallero, Zamora, U2D2}. The effective scattering length in the plane $x-y$ becomes $a_{\sigma\sigma'} \rightarrow a_{\sigma\sigma'}/\sqrt{2 \pi} l_z$, with $l_z=\sqrt{\hbar/m \omega_z}$, being $\omega_z$ a typical frequency of condensates confined in 2D \cite{Hadzibabic, Lung-Hung}. In current experiments the values of the coupling constants $g_{\sigma\sigma'}$, can be changed via Feshbach resonances, and thus tuned to have either, equal or different values of the intra and inter-species interactions, that is $g_{\uparrow \uparrow}= g_{\downarrow \downarrow}= g_{\uparrow \downarrow}$, or $g_{\uparrow \uparrow}=g_{\downarrow \downarrow} \neq g_{\uparrow \downarrow}$. In the present investigation we consider $g_{\uparrow \uparrow}= g_{\downarrow \downarrow}$ and $g_{\uparrow \downarrow}= g_{\downarrow \uparrow}$. As stated in \cite{Wang, Papp}, miscibility of the two component mixture is determined by the relation between $g_{\uparrow \uparrow}$, $g_{\downarrow \downarrow}$ and $g_{\uparrow \downarrow}$, as a matter of fact, the separation of the hyperfine components happens when the condition $g_{\uparrow \downarrow} > \sqrt{g_{\uparrow \uparrow} g_{\downarrow \downarrow}}$ is satisfied. In our analysis we shall consider a ratio of the intra and inter-species interaction that guaranty miscibility of the hyperfine components, the specific value that we use is $g_{\uparrow \downarrow}/g_{\uparrow \uparrow}$=0.8. 

The potential depth is scaled in units of the recoil energy $E_R= \frac{\hbar^2 k^2}{2m}$, where $k=\pi / a_0$ and $a_0$ being the lattice constant of the original square and hexagonal lattices. As previously shown in the literature, the mean-field approximation describes well the dynamics of initial magnetic domains, that evolve in time under given conditions  \cite{Ray, Schulte, Adhikari, Michelangeli, CMadronero}.

\subsection{Preparation of the initial ferromagnetic state}
\label{Initialstates}
\begin{figure}[h]
\begin{center}
\includegraphics[width=8.5cm, height=8.5cm]{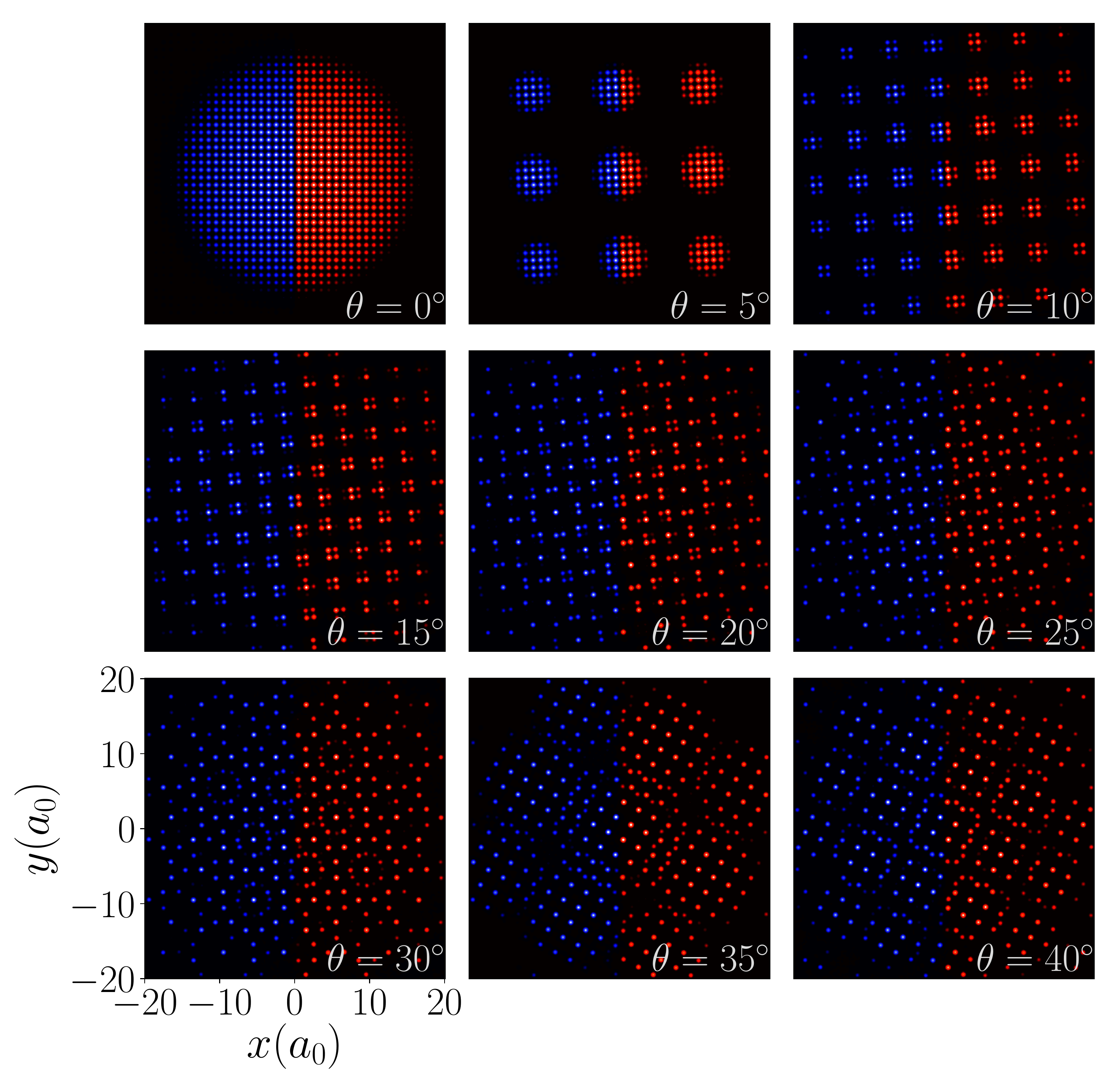}
\end{center}
\caption{Density profiles prepared from the superposition of two rotated square lattices a twisting angle $\theta$. Left (blue) and right (red) sides correspond to the superfluid density associated to the pseudo-spin components $\downarrow$ and $\uparrow$. Frequency of the harmonic trap is $\omega_r=0.4 \omega_0$.}
\label{Figure1}
\end{figure}

\begin{figure}[h]
\begin{center}
\includegraphics[width=8.5cm, height=5.7cm]{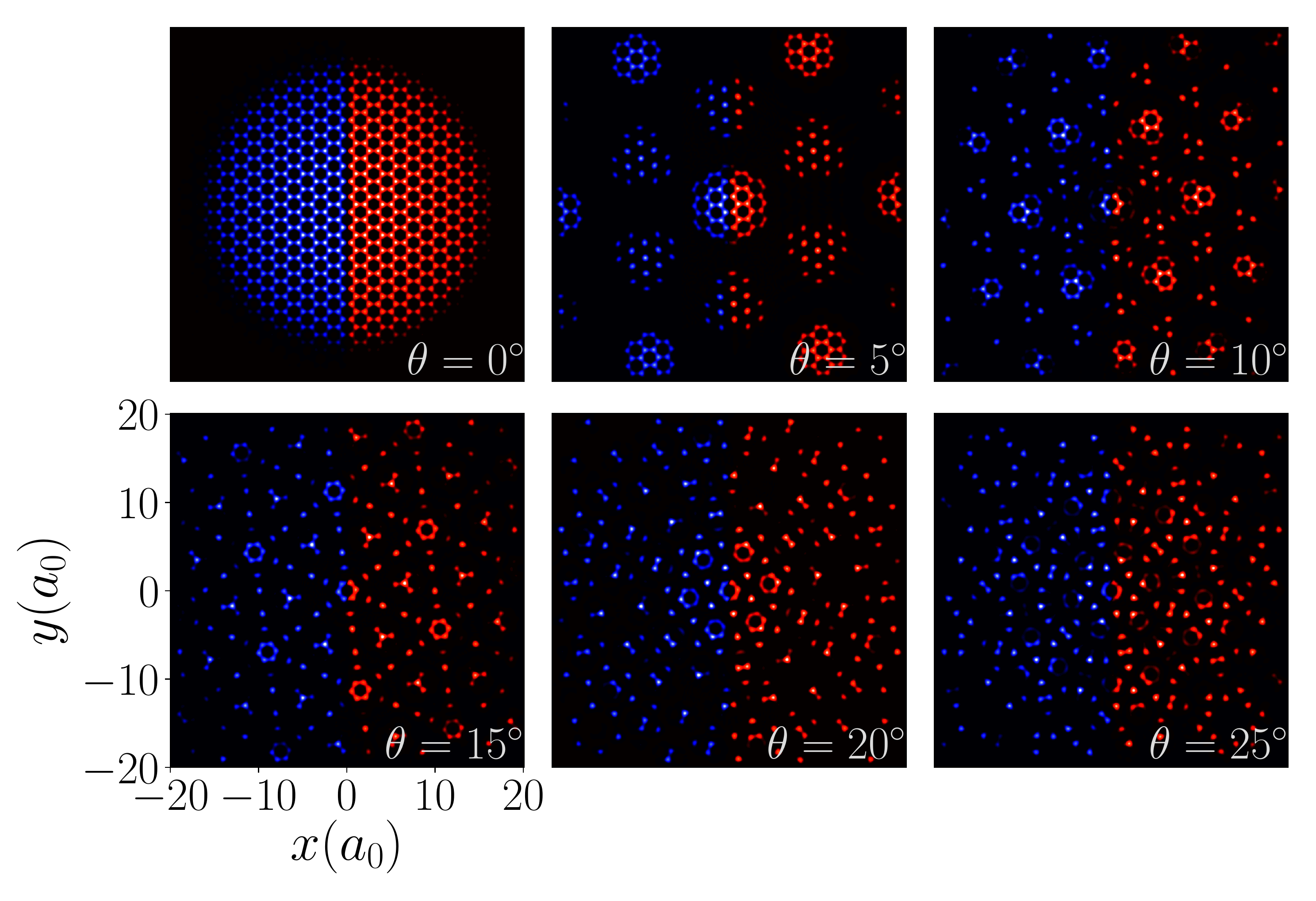}
\end{center}
\caption{Density profiles prepared from the superposition of two rotated hexagonal lattices a twisting angle $\theta$. Left (blue) and right (red) sides correspond to the superfluid density associated to the pseudo-spin components $\downarrow$ and $\uparrow$. Frequency of the harmonic trap is $\omega_r=0.4 \omega_0$.}
\label{Figure2}
\end{figure}

The initial state associated to a given angle $\theta$ for which a double ferromagnetic state lies in a moir\'e lattice, either square or hexagonal, is set as we describe in the next lines. First we determine the stationary state for the coupled equations (\ref{coupledGP}) for optical lattices defined by the potential $V_{ \mathrm{ext}}(x,y)$, having a constant depth $V_{0}/E_R = 4$. For this purpose free energy minimization is performed by means of imaginary time evolution  $\tau\rightarrow it$  \cite{Calculations,Dum,Zhang}. After this procedure, we manually remove the particles having spin component $\sigma =\uparrow$ from the left half layer, while particles with $\sigma=\downarrow$ from the right half layer. In Figs. \ref{Figure1} and \ref{Figure2} we plot some of the density profiles prepared as the initial state for the square and hexagonal moir\'e structures respectively. The frequency of the harmonic trap in these figures is $\omega_r=0.4 \omega_0$. At the bottom of each subfigure it is indicated the twisting angle. Blue and red colors identify the density profiles of the $\downarrow$ and $\uparrow$ components. Notice that equal ground state densities of different hyperfine states remain at each half in the 2D moir\'e lattices. The removal of particles proposed to set these initial states mimics experimental procedures in which a digital mirror device is used to optically remove the particles at specific positions \cite{Choi,Takahashi}. Certainly, another route to achieve ferromagnetic domains is by means of a magnetic field \cite{Weld}.  These patterns manually created, that is the two ferromagnetic domains for a given value of the twisting angle $\theta$, are our starting point to study its time evolution under the influence of the moir\'e confinement. We should note here that the initial state prepared for each angle $\theta$ is non-stationary, and consequently it evolves under their own dynamics. We must point that this kind of states from which a system evolves under its own dynamics are the so called quantum quenches created in the laboratory. Our particular interest is to investigate how the local magnetization of the ferromagnetic domains degrades when the weakly interacting 2D Bose mixture evolves in the absence of other external fields, except the one produced from the superposition of the lattices forming the moir\'e patterns associated to either, square or hexagonal  moir\'e patterns. It is important to mention that the effective coupling interaction coefficients must be rescaled for $t>0$ since half of the population is removed to have the magnetic domains, that is $N \rightarrow N/2$.  

Due to the symmetry of the lattices, the twisting angles that shall be considered for the analysis of the evolution in time are $\theta \in (0^\circ,30^\circ]$ and $\theta \in (0^\circ,45^\circ]$ for the hexagonal and square lattices respectively. Angles larger than those produce the same results because of the mirror symmetry. 

\subsection{Square and hexagonal moir\'e crystals}
\label{crystals}

\begin{figure}[h]
\begin{center}
\includegraphics[width=8.5cm, height=4.5cm]{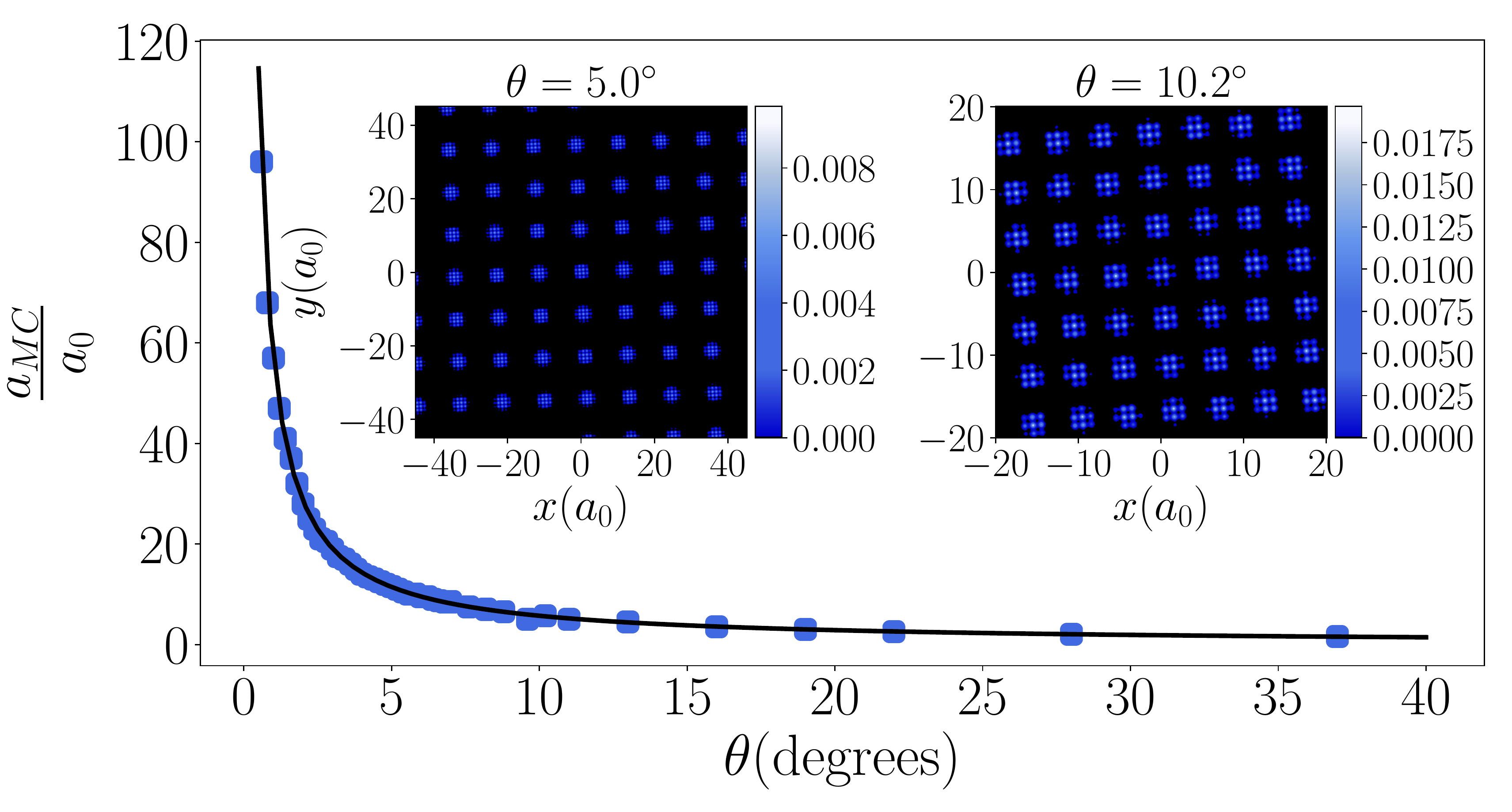}
\end{center}
\caption{Lattice constant $a_{MC}$ of square moir\'e crystals as a function of the twisting angle $\theta$, $a_0$ is the lattice constant for $\theta=0^\circ$. Dots in this curve are fitted by the formula in Eq. \ref{fit_aMC} (solid line). The insets show the superfluid density of the stationary state for one of the spin components for two different rotation angles. Blue color scheme in the bar indicate the size of the superfluid density. Left and right sides correspond to $\theta=5^\circ$ and $10.2^\circ$ respectively.}
\label{Figure3}
\end{figure}

\begin{figure}[h]
\begin{center}
\includegraphics[width=8.5cm, height=4.5cm]{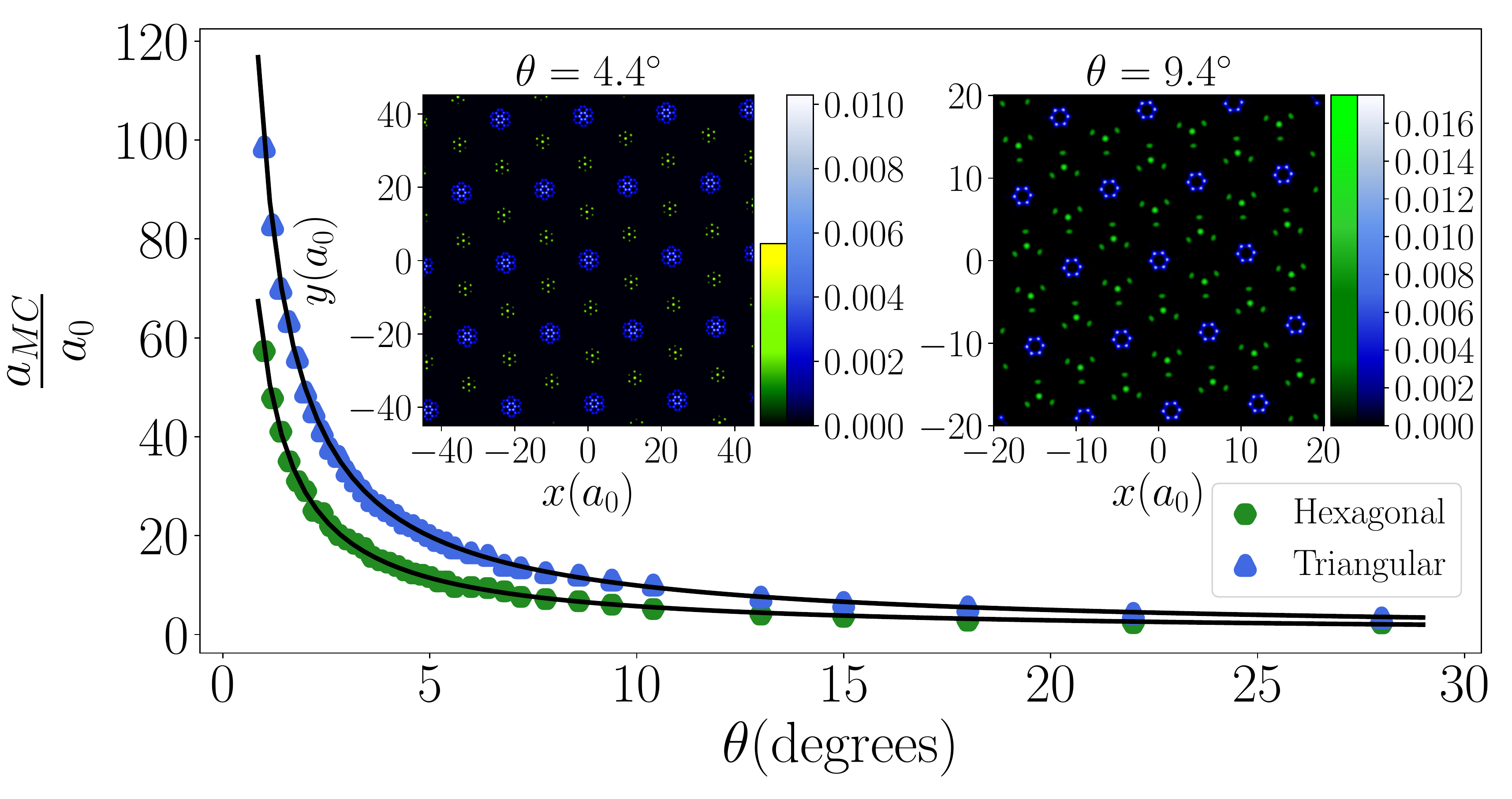}
\end{center}
\caption{Lattice constant $a_{MC}$ of hexagonal moir\'e crystals as a function of the twisting angle $\theta$, $a_0$ is the lattice constant for $\theta=0^\circ$. Blue and green dots in the curves are fitted by the formula in Eq. \ref{fit_aMC} (solid lines) and correspond to two superimposed triangular and hexagonal lattices, that emerge when two hexagonal lattices are rotated an angle $\theta$. The insets show the superfluid density of the stationary state for one of the spin components for two different rotation angles. Blue and green density color schemes in the bars indicate the size of the superfluid density. Left and right sides correspond to $\theta=4.4^\circ$ and $9.4^\circ$ respectively.}
\label{Figure4}
\end{figure}

Before presenting the discussion of the dynamics of the double magnetic domain, we must point out some facts associated with the stationary moir\'e patterns used to prepare the initial states. The large size of the lattices considered in our study, namely $90 \times 90$ sites, allowed us to establish several conclusions regarding the twisting angle and the patterns emerging from the superposition of the lattices. From the stationary density profiles of each component, for a frequency $\omega_r=0$, we observe that starting at low angles of relative rotation, the so called moir\'e crystals appear \cite{Feuerbacher, Zengming}. These moir\'e crystals, also called super lattices, are characterized by having a lattice constant that depends on the twisting angle $\theta$. We observe that such lattices are present for twisting angles in the interval $\theta \in \left(0^\circ ,30^\circ \right)$ and $\theta \in \left(0^\circ ,45^\circ \right)$ for hexagonal and square geometries respectively. In Figs. \ref{Figure3} and \ref{Figure4} we plot the lattice constan $a_{MC}$ as a function of the angle $\theta$. The lattice constant of the super lattice is measured in units of the lattice parameter $a_0$ that characterizes the original square and hexagonal structures.
To illustrate the dependence of the lattice constant $a_{MC}$ on the twisting angle $\theta$, in Figs. \ref{Figure3} and \ref{Figure4}, besides the behavior of $a_{MC}$ vs. $\theta$, we show in the insets a couple of density profiles for the single component $\Psi_{\uparrow}$ associated with two different values of $\theta$. The amplitude of the such profiles is shown in a density color scheme scaled with the bar at the right of the insets. As one can see from these profiles, instead of observing sharply defined spots of the superfluid density, we observe a basis o pattern at each {\it node} of the super lattice. We notice how these patterns inherit the symmetry of the original square and hexagonal structures.  
It is important to point out that in the case of moir\'e lattices arising from hexagonal patterns, two superlattices emerge as a result of the relative rotation between the original lattices. These lattices having hexagonal and triangular symmetries are such that, for some values of the twisting angle, one of them has a density profile with a larger intensity with respect to the other (see in Fig. \ref{Figure4} the color density scheme at the bar on the right of each profile). Blue and green colors correspond to triangular and hexagonal moir\'e crystals. 

The formulae that fits the dependence of $a_{MC}$ as a function of $\theta$ for the superlattices is \cite{Feuerbacher, Zengming}, 
\begin{equation}
a_{MC}=\frac{a_0}{na \sin (\theta/bn)},
\label{fit_aMC}
\end{equation}  
where $n$ is a non-zero integer number, and $a=b=1$ for the square and hexagonal moir\'e crystals, while $a=1.56$ and $b=2.70$ for the triangular moir\'e crystal.

\begin{figure}[h]
\begin{center}
\includegraphics[width=8.0cm, height=4.4cm]{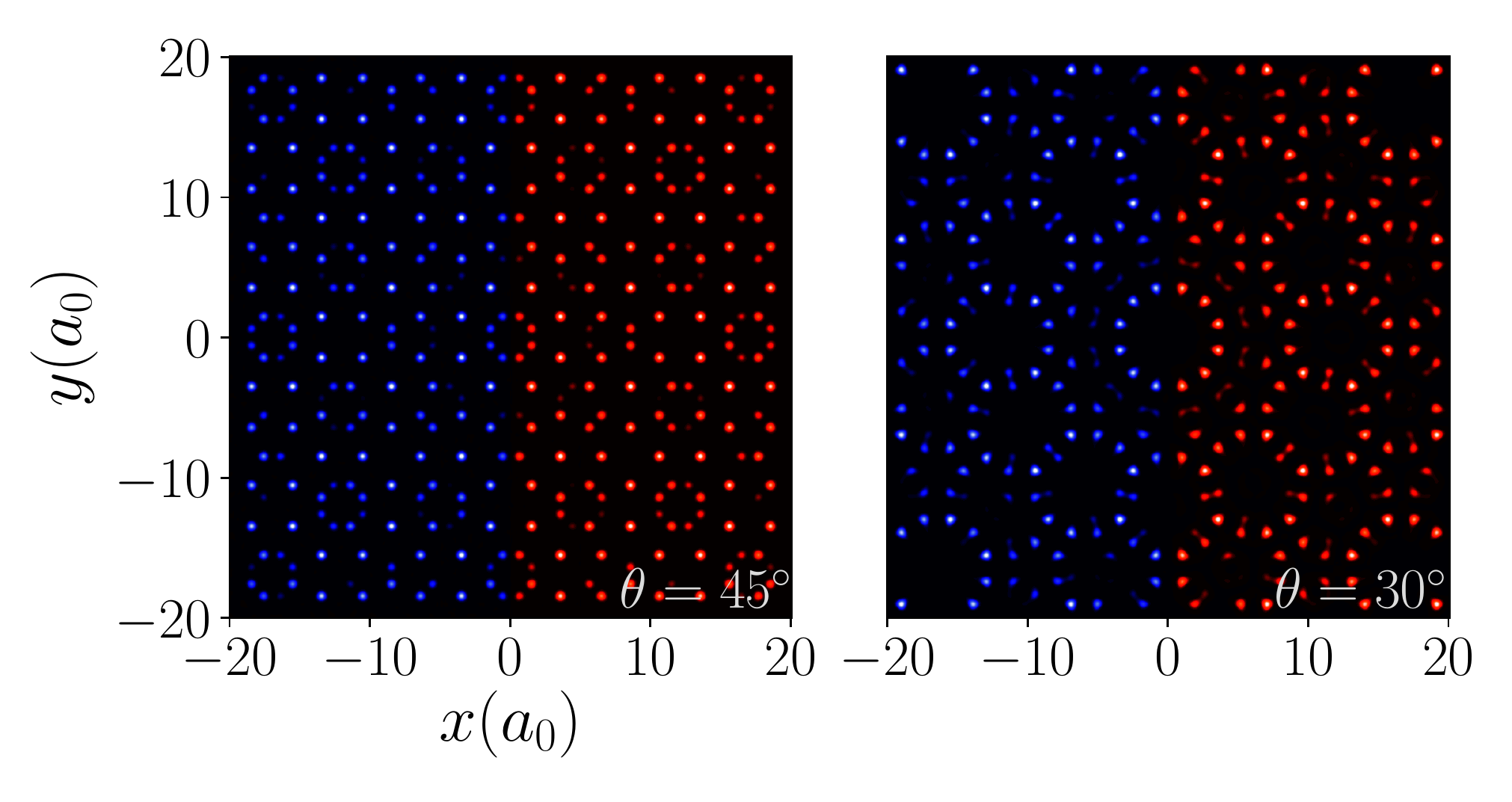}
\end{center}
\caption{Density profiles prepared from the superposition of two rotated square (left) and two rotated hexagonal (right) lattices, being the rotation angles $\theta =45 ^\circ$ and $\theta =30^\circ$ respectively. For those angles the resulting structures are quasicrystals with octagonal and hexagonal rotation symmetry. Left (blue) and right (red) sides in each panel correspond to the superfluid density associated with the pseudo-spin components $\downarrow$ and $\uparrow$ respectively. Frequency of the harmonic trap is $\omega_r=0$.}
\label{Figure5}
\end{figure}

As expected, the moir\'e patterns that result from two superimposed square and hexagonal structures, for the particular angles $\theta =45 ^\circ$ and $\theta =30^\circ$, have octagonal and hexagonal quasicrystalline geometries respectively. In Fig. \ref{Figure5} we illustrate these quasicrystalline structures associated to $\uparrow$ and $\downarrow$ components for the case in which the harmonic confinement is absent. Those quasicrystalline moir\'e patterns shall also be considered as initial state, to follow its time evolution.

\section{Dynamics of a double magnetic domain in Moir\'e lattices}
\label{dynamics}

The physical observable that accounts for the magnetic character of the system here analyzed is the spin texture. This quantity is defined as,

\begin{equation}
\begin{split} 
\textbf{T}(x,y,t)=\Psi^{\dagger}(x,y,t)\textbf{F}\Psi(x,y,t),
\end{split}
\end{equation}
where the operator $\textbf{F}$ is written in terms of the Pauli matrices $\textbf{F}=(\sigma_x,\sigma_y,\sigma_z)$, and $\Psi(x,y,t)=(\Psi_{\uparrow}(x,y,t),\Psi_{\downarrow}(x,y,t))$. Thus the spin texture, which a real quantity, is given by,
\begin{equation}
\begin{split} 
\textbf{T}(x,y,t)=&\left[\Psi_{\uparrow}^{*}\Psi_{\downarrow}+\Psi_{\uparrow}\Psi_{-\downarrow}^{*}\right]\hat{x}\\
&+i\left[\Psi_{\uparrow}^{*}\Psi_{\downarrow}-\Psi_{\uparrow}\Psi_{\downarrow}^{*}\right]\hat{y} \\
&+\left[{\left|{\Psi}_{\uparrow}\right|}^{2}-{\left|{\Psi}_{\downarrow}\right|}^{2}\right]\hat{z}.
\end{split}
\label{Textura}
\end{equation}

The component along axis $\hat{z}$ is the local magnetization at time $t$. This is precisely the observable that we shall track to determine in a quantitative way either, the persistence or the absence of the initial magnetic domains as a function of time, for a given value of the twisting angle $\theta$. In addition, we must emphasize that this observable is the natural quantity that can be accessed in typical experiments with ultracold atoms \cite{Palacios}, or within the context of condensed matter.

The observables to be studied are the magnetizations $m_L$ and $m_R$, in the left and right sides of the lattice, as a function of time for different values of $\theta$, and as stated above, at a fixed value of the coupling interactions. These quantities are defined in terms of the local magnetization $m(x,y;t) = \rho_{\uparrow}(x,y;t)-\rho_{\downarrow}(x,y;t)$, where $\rho_{\uparrow}(x,y;t)$ and $\rho_{\downarrow}(x,y;t)$ are the densities associated with the components $\uparrow$ and $\downarrow$ respectively. Thus, magnetization in left and right sides are,
\begin{eqnarray}
m_L=\int  \int_{\Omega_L} dx \ dy \> m(x,y;t)\cr
m_R=\int  \int_{\Omega_R} dx \ dy \> m(x,y;t),
\end{eqnarray}
where $\Omega_L$ and $\Omega_R$ are the left and right halves of the system, respectively. Because of the particular election of the initial state we have that $m_L(t=0) =-0.5$ and $m_R(t=0)=0.5$.

The evolution in time of the initial states will be followed in dimensionless time $\tau=E_R t/ \hbar$. It is important to mention here that all of our numerical calculations were performed ensuring that changing $\tau \rightarrow -\tau$, at any temporal step along the time dynamics, allows us to recover the initial state. The time during which the magnetization in left and right sides will be tracked, coincides with that at which the magnetization in both sides becomes null for a twisting angle $\theta=0^\circ$. In other words, the magnetization associated with $\theta=0 ^\circ$ when no moir\'e patterns exist, provides a reference in terms of which one can evaluate the influence of changing the angle $\theta$ in either, diminishing or preserving the magnetization. As we describe below, the presence of the harmonic confinement also plays a role in the evolution of the initial magnetization.
\begin{figure}[h]
\begin{center}
\includegraphics[width=8.3cm, height=8.3cm]{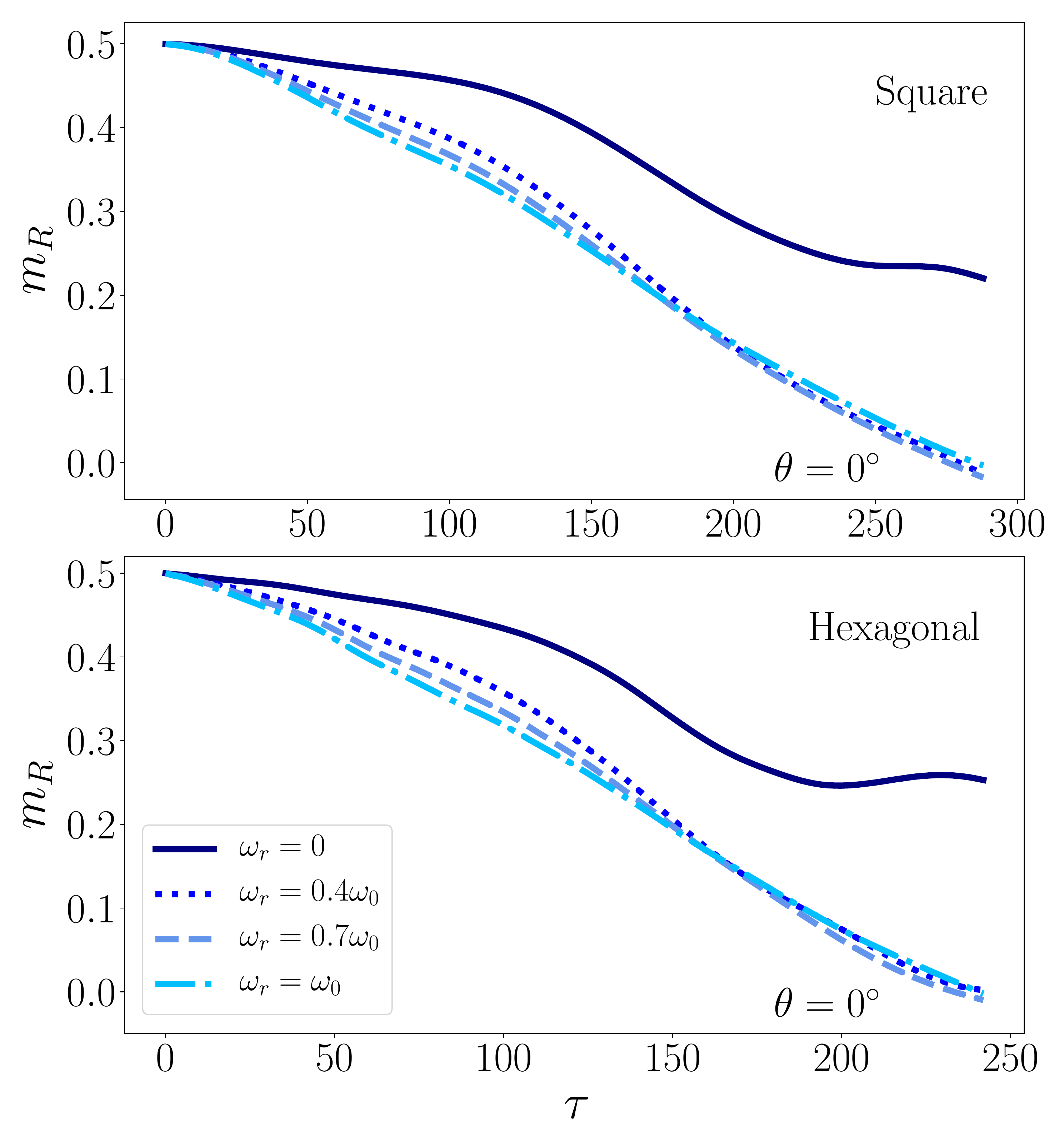}
\end{center}
\caption{Magnetization as a function of time in the right side in square (top panel) and hexagonal (bottom panel) lattices. Different curves in both panels are associated with the size of the harmonic confinement as indicated in the bottom panel. }
\label{Fig6}
\end{figure} 

In Fig. \ref{Fig6} top panel and bottom panel we plot the magnetization on the right side for $\theta=0^\circ$, for square and hexagonal structures respectively. One can appreciate from these figures the influence of the harmonic confinement in the demagnetization process. While for non-zero values of the ratio $\omega_r/\omega_0$ the initial domains are destroyed around $\tau=300$, in the homogeneous case, that is $\omega_r/\omega_0=0$, magnetization in left and right sides is retained for longer times, $\tau \sim 700$ and $\tau \sim 600$ for square and hexagonal lattices respectively (not visible in the figure). This behavior can be understood as follows. In the absence of harmonic confinement the hyperfine spin components evolve under the influence of both, the effective mean field interactions, and the original square or hexagonal lattices, without any other additional field controlling the dynamics. In contrast, when the harmonic confinement is turned on, the spin components placed on left and right sides will experience a field directed towards the center of the trap, that is, towards the origin. Associated with the effective field experienced by each component, they will tend to occupy the same space, and therefore they will mix, thus destroying the initial magnetic order in left and right sides. As the size of the frequency is increased, the annihilation of the initial domain become more visible. However, as we explain in the next paragraphs, the scenario completely changes when the two superimposed lattices are rotated one with respect to each other.    

Taking as a reference the behavior of the magnetization for $\theta=0^\circ$, we proceed with the analysis of the magnetization dynamics as a function of $\theta$ for the square and hexagonal moir\'e patterns, considering also the influence of the harmonic confinement. The main conclusion found for the time evolution of the initial bi-component magnetic domain is the identification of an interval of the twisting angles for which the initial state is preserved, disregarding the value of the ratio $\omega_r/\omega_0$. 
\begin{figure}[h]
\begin{center}
\includegraphics[width=8.3cm, height=8.3cm]{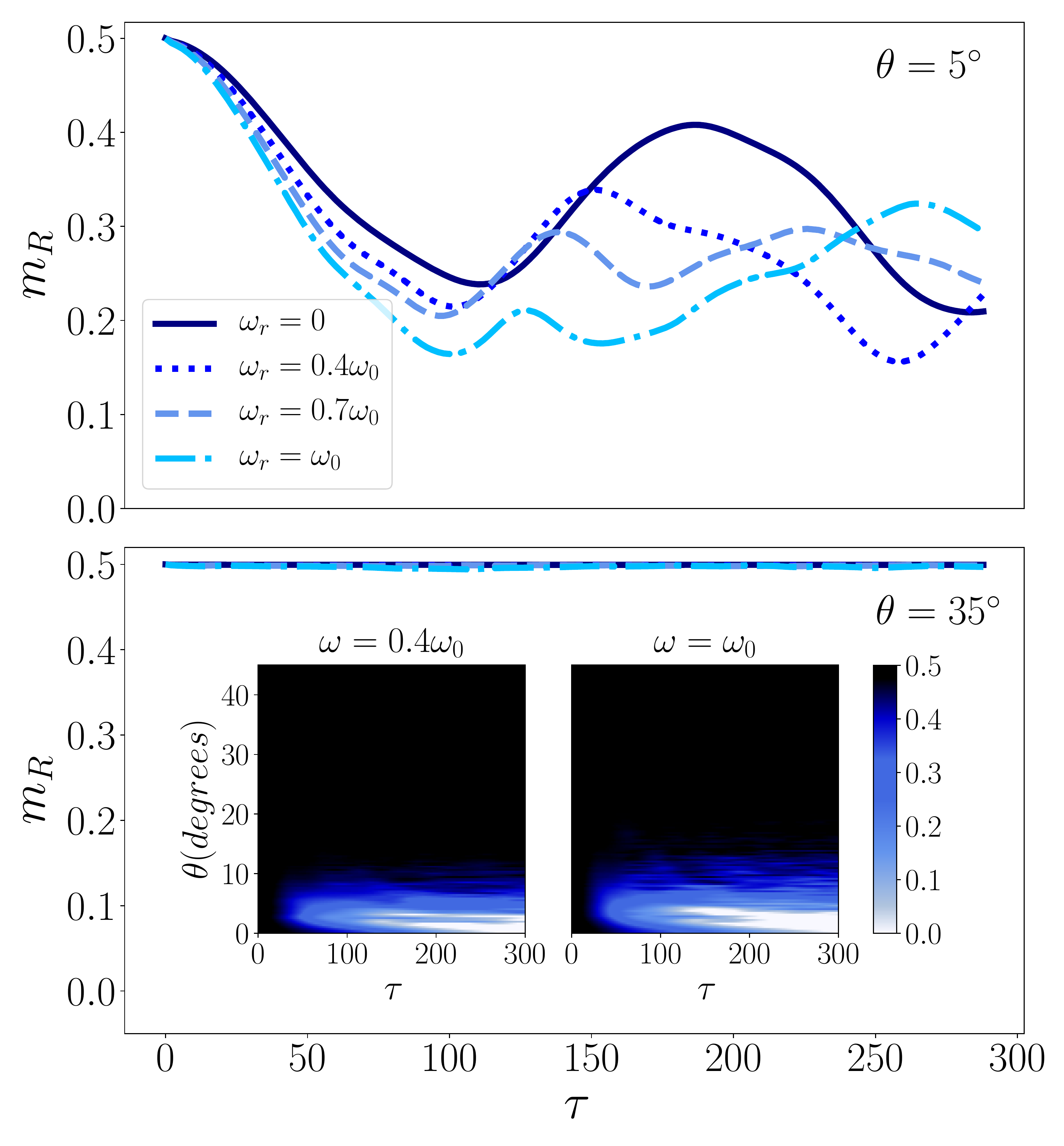}
\end{center}
\caption{Magnetization as a function of time in the right side of a moir\'e patterns created from the superposition of two square lattices rotated an angle $\theta$. Panels on top and bottom correspond to $\theta=5 ^\circ$ and $\theta=35 ^\circ$ respectively. Different lines in both panels are associated with the size of the harmonic confinement as indicated in the top panel. The insets at the bottom figure are the value of the magnetization in the right side as a function of time for the whole range of angles considered in our analysis.}
\label{Fig7}
\end{figure}
In Figs. \ref{Fig7} and \ref{Fig8} we plot the magnetization on the right side for square and hexagonal moir\'e patterns respectively, for two different values of the twisting angle $\theta$. The angles were chosen to illustrate the dependence of the behavior of $m_R(\tau)$ on the moir\'e structures generated from the rotation of the pair of lattices for four different values of the frequency, $\omega_r=$ 0, $0.4 \omega_0$, $0.7 \omega_0$ and $\omega_0$. The behavior as a function of time for different values of these frequencies is identified with solid, dotted, dashed, and dashed-dotted lines respectively. As one can see from top panels of figures \ref{Fig7} and \ref{Fig8}, for $\theta =5^\circ$ the initial magnetization rapidly get lost disregarding the presence of the harmonic confinement, and, although $m_R$ certainly oscillates as a function of time, no longer recover its initial value. In contrast, we appreciate from bottom panels of figures \ref{Fig7} and \ref{Fig8} that $m_R$ shows no variations when $\theta=35 ^\circ$ and $\theta=30 ^\circ$ respectively. To better illustrate the influence that the moir\'e structures here considered have on preserving the initial state, we followed the dynamics of $m_R(\tau)$ in the whole interval of rotation angles that give rise the formation of square and hexagonal moir\'e patterns. In the insets of figures \ref{Fig7} and \ref{Fig8} we plotted the behavior of $m_R(\tau)$ for $\omega= 0.4 \omega_0$ (left) and $\omega=\omega_0$ (right). The insets show in a density color scheme the variations of $m_R(\tau)$ for $\theta \in [0^\circ,45^\circ]$ and $\theta \in [0^\circ,30^\circ]$ for square and hexagonal lattices respectively. We observe from these figures that moir\'e patterns arising from rotation angles larger than $\theta \gtrsim10 ^\circ$ constitute structures in which the magnetization is preserved.

\begin{figure}[h]
\begin{center}
\includegraphics[width=8.3cm, height=8.3cm]{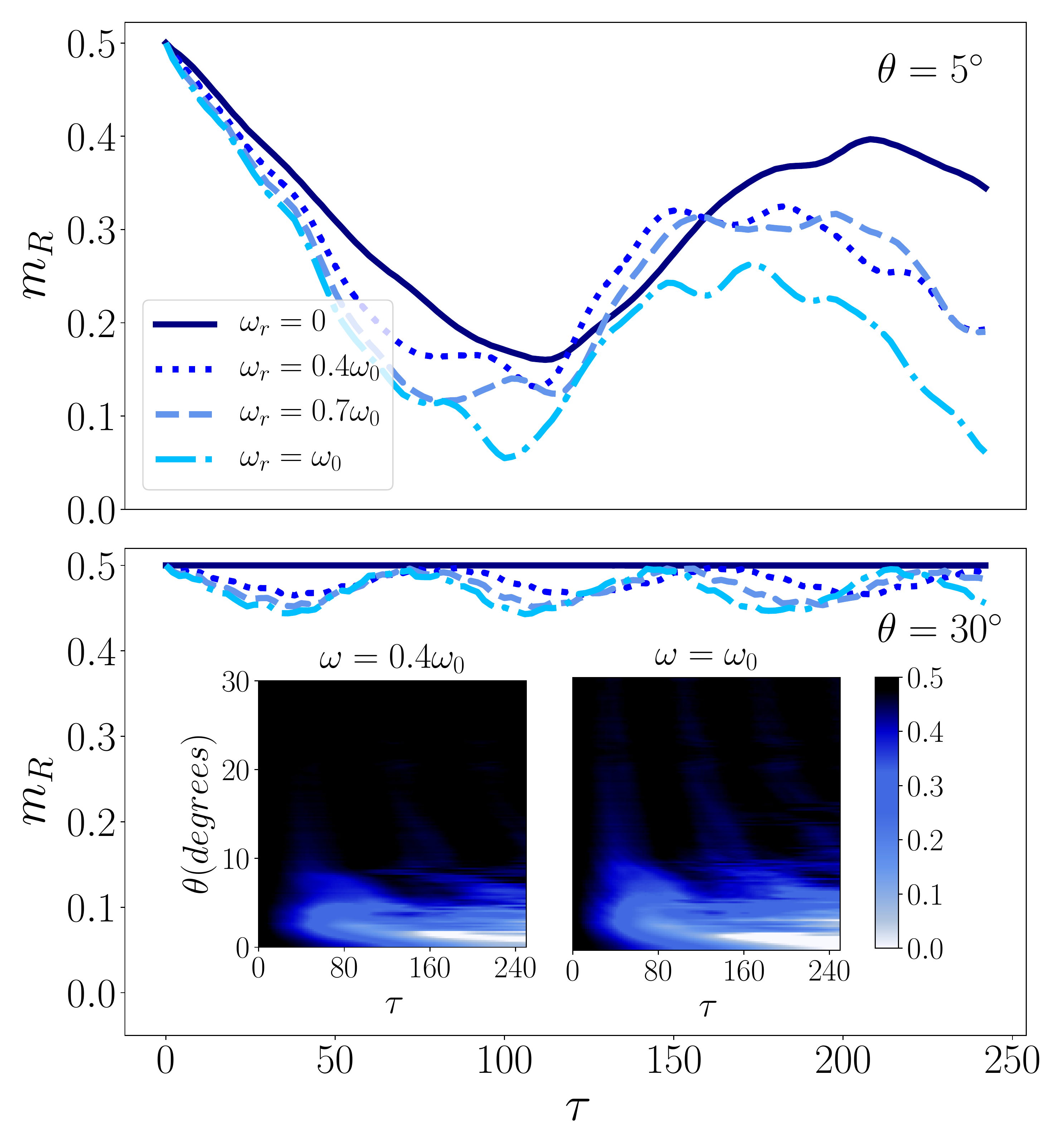}
\end{center}
\caption{Magnetization as a function of time in the right side of a moir\'e patterns created from the superposition of two hexagonal lattices rotated an angle $\theta$. Panels on top and bottom correspond to $\theta=5 ^\circ$ and $\theta=30 ^\circ$ respectively. Different lines in both panels are associated with the size of the harmonic confinement as indicated in the top panel. The insets at the bottom figure are the value of the magnetization in the right side as a function of time for the whole range of angles considered in our analysis.}
\label{Fig8}
\end{figure}

\section{Conclusions}
\label{Conclusion}
We have investigated the stationary states and the time dynamics of a weakly interacting Bose gas lying in moir\'e patterns with square or honeycomb symmetries.  The patterns considered arose from the superposition of two square or hexagonal lattices rotated by an angle $\theta$, plus the presence of an isotropic harmonic confinement in 2D. The system was described within the mean field approach through the time dependent Gross-Pitaevskii equation that was solved for lattices with a number of sites up to $\sim 90 \times 90$. For the analysis of the stationary states we considered a single hyperfine spin component of the Bose condensate, and performed numerical calculations that allowed us to recognize the continuous set of crystalline lattices were the superfluid can be placed. These lattices known as moir\'e crystals or superlattices, emerge for twisting angles belonging to the intervals $\theta \in \left(0^\circ ,30^\circ \right)$ and $\theta \in \left(0^\circ ,45^\circ \right)$ for hexagonal and square geometries respectively. We found that the lattice parameter of the moir\'e crystals $a_{MC}$ scales with the lattice constant $a_0$ of the original lattices as $a_{MC}=\frac{a_0}{na \sin (\theta/bn)}$, with $n \in \mathbb{N}$ and $a$ and $b$ constant numbers. For rotation angles $\theta =45^\circ$ and $30^\circ$ between two square and hexagonal lattices respectively the superfluid component remains arranged in octagonal and hexagonal quasicrystalline structures. 

For the dynamics we considered the simplest bi-component magnetic domain lying in moir\'e patterns and followed their time evolution, with the interest of establishing their prevalence as a function of the angle that define the particular moir\'e lattice. The initial state was prepared by situating two different hyperfine spin components of the condensate on the left and right sides of a given moir\'e lattice, and then leaved to evolve under the influence of the effective constant interactions, and the potential created from the superposition of the rotated lattices and the isotropic harmonic confinement. The main result obtained by tracking the evolution of the spin texture, and particularly the magnetization on left and right sides, was the identification of an interval of the twisting angles $\theta \in [0^\circ,10^\circ]$ for which the initial state deteriorates for both square and hexagonal moir\'e lattices. In other words, we found that a double magnetic domain placed on moir\'e lattices remains in time whenever the angle that defines the moir\'e pattern, either with square or hexagonal symmetry, exceeds $\theta>10^\circ$. 

The investigation here presented adds to the understanding of magnetic properties that arise in van der Waals heterostructures, and particularly to the study of magnetic domains, either, created spontaneously or arranged by means of external fields, in superfluid, excitons, and electrons, confined in moir\'e lattices. At the same time that the proposal here presented uses moir\'e lattices as a quantum simulator \cite{Kennes}, it employs the ultracold quantum gases as the ideal platform to set and probe the dynamics of magnetic domains in 2D.  Our analysis is relevant within the current context of the design of quantum materials belonging to the emerging field of twistronics, and/or the possibility of engineering quantum simulation platforms to create quantum memories.

\acknowledgments{
This work was partially funded by Grant No. IN108620 from DGAPA (UNAM). C.J.M.C acknowledges CONACYT scholarship.
}

\appendix
\section{Numerical Calculations Details}
The results here reported were obtained from numerical solution of the coupled GP equations. In this appendix we present additional details related to the technical part of the calculations performed. To find the solution of the Eq. \ref{coupledGP} we use the fourth-order Runge-Kutta method, which we used both, to obtain the ground state and to study the dynamics of the system. The numerical parameters used for the simulations are shown in table \ref{Table1}, while the physical parameters of the system are presented in table \ref{Table2}.

\begin{table}[h]
\centering
\small
\begin{tabular}{p{4.5cm} c  c} 
Name & Symbol & Value\\
\hline\hline
Number of grid points in the $x$ direction & $N_{x}$ & 512-1024\\
Number of grid points in the $y$ direction & $N_{y}$ & 512-1024\\
Spatial extension of the numerical grid in the $x$ direction & $L_{x}$ & 40-90 $a_0$ \\
Spatial extension of the numerical grid in the $y$ direction  & $L_{y}$ & 40-90 $a_0$ \\
Step size used in real time evolution & $d\tau$ & $0.001$\\
\end{tabular}
\caption{Parameters for the numerical simulation}
\label{Table1}
\end{table}

\begin{table}[h]
\centering
\small
\begin{tabular}{p{4.5cm} c  c} 
Name & Symbol & Value\\
\hline\hline
Particle number & $N$ & 600 \\
$^{87}$Rb mass & $m$ & $86.9$ amu\\
Lattice constant & $a_0$ & $532$ nm\\
Reference trap frequency & $\omega_{0}$ & $2\pi\times 50$ rad/s\\
Trap frequency $(z)$ & $\omega_{z}$ & $2\pi\times 5000$ rad/s\\
Bare $s$-wave scattering length  & $a_{\uparrow \uparrow}$=$a_{\downarrow \downarrow}$ & 100.4 $a_{B}$ \\
Potential depth  & $V_{0}$ & 4 $E_R$  \\
\end{tabular}
\caption{Physical parameters used in the numerical simulation}
\label{Table2}
\end{table}


\begin{thebibliography}{9}

\bibitem{WangY} Y.Wang, Z. Wang, W. Yao, G-B Liu, and H. Yu \href{} {Phys. Rev. B {\bf 95}, 115428 (2017).}

\bibitem{Ponomarenko} L. A. Ponomarenko, R. V. Gorbachev, G. L. Yu, D. C. Elias, R. Jalil, A. A. Patel, A. Mishchenko, A. S. Mayorov, C. R. Woods, J. R. Wallbank, M. Mucha-Kruczynski, B. A. Piot, M. Potemski, I. V. Grigorieva, K. S. Novoselov, F. Guinea, V. I. Fal'ko, and A. K. Geim, \href{https://www.nature.com/articles/nature12187} {Nature  {\bf 497}, 594 (2013).}

\bibitem{Dean} C. R. Dean, L. Wang, P. Maher, C. Forsythe, F. Ghahari, Y. Gao, J. Katoch, M. Ishigami, P. Moon, M. Koshino, T. Taniguchi, K. Watanabe, K. L. Shepard, J. Hone, and P. Kim, \href{https://www.nature.com/articles/nature12186} {Nature {\bf 497}, 598 (2013).}

\bibitem{Hunt} B. Hunt, J. D. Sanchez-Yamagishi, A. F. Young, M. Yankowitz, B. J. LeRoy, K. Watanabe, T. Taniguchi, P. Moon, M. Koshino, P. Jarillo-Herrero, and R. C. Ashoori, \href{https://www.science.org/doi/10.1126/science.1237240} {Science {\bf 340}, 1427 (2013).}

\bibitem{Cao} Y. Cao, V. Fatemi, S. Fang, K. Watanabe, T. Taniguchi, E. Kaxiras and P. Jarillo-Herrero, \href{https://www.nature.com/articles/nature26160}, {Nature {\bf 556}, 43 (2018).}

\bibitem{Cao2} Y. Cao, V. Fatemi, A. Demir, S. Fang, S. L. Tomarken, J. Y. Luo, J. D. Sanchez-Yamagishi, K. Watanabe, T. Taniguchi, E. Kaxiras, R. C. Ashoori  and P. Jarillo-Herrero,  \href{https://www.nature.com/articles/nature26154} {Nature (London) {\bf 556}, 80 (2018)}.

\bibitem{Xiao} F. Xiao, K. Chen, and Q. Tong,  \href{https://journals.aps.org/prresearch/abstract/10.1103/PhysRevResearch.3.013027}, {Phys. Rev. Res. {\bf 3}, 013027 (2021).}

\bibitem{Tarnopolsky} G. Tarnopolsky, A. J. Kruchkov and  A. Vishwanath, \href{https://journals.aps.org/prl/abstract/10.1103/PhysRevLett.122.106405} {Phys. Rev. Lett. {\bf 122}, 106405 (2019).}

\bibitem{Yndurain} F. Yndurain, \href{https://journals.aps.org/prb/abstract/10.1103/PhysRevB.99.045423} {Phys. Rev. B {\bf 99}, 045423 (2019)}.

\bibitem{Lopez-Bezanilla} A. Lopez-Bezanilla, \href{https://journals.aps.org/prmaterials/abstract/10.1103/PhysRevMaterials.3.054003} {Phys. Rev. Mater. {\bf 3}, 054003 (2019).}

\bibitem{Kangjun} K. Seo, V. N. Kotov, and B. Uchoa, \href{https://journals.aps.org/prl/abstract/10.1103/PhysRevLett.122.246402}, {Phys. Rev. Lett. \textbf{122}, 246402 (2019)}.

\bibitem{Pan} H. Pan, F. Wu, and S.Das Sarma \href{Phys. Rev. B  102, 201104 (2020)}, {Phys. Rev. B {\bf 102}, 201104 (2020).}

\bibitem{GTudela} A. Gonz\'alez-Tudela and J. I. Cirac, \href{https://journals.aps.org/pra/abstract/10.1103/PhysRevA.100.053604}, {Phys. Rev. A {\bf 100}, 053604 (2019).}

\bibitem{Zengming} Z. Meng, L. Wang, W. Han, F. Liu, K. Wen, Ch. Gao, Ch. Chinand J. Zhang \href{https://arxiv.org/abs/2110.00149},{arXiv:2110.00149 (2021).}

\bibitem{Mahood} R. Mahmood, A. Vela Ramirez, and A. C. Hillier, \href{https://pubs.acs.org/doi/pdf/10.1021/acsanm.1c00210}, {ACS Appl. Nano Mater. \textbf{4}, 9, 8851 (2021).} 

\bibitem{Choi} J.-Y. Choi, S. Hild, J. Zeiher, P. Schau$\beta$, A. Rubio-Abadal, T. Yefsah, V. Khemani, D. A. Huse, I. Bloch, and C. Gross, \href{https://www.science.org/doi/10.1126/science.aaf8834}, {Science {\bf 352}, 1547 (2016)}.

\bibitem{Takahashi} F. Sch\"afer , T. Fukuhara , S. Sugawa , Y. Takasu and Y. Takahashi, \href{https://www.nature.com/articles/s42254-020-0195-3}, {Nat. Rev. Phys. {\bf 2}, 411 (2020).} 

\bibitem{Hadzibabic} Z. Z. Hadzibabic, P. Kr\"uger, M. Cheneau, B. Battelier and J. Dalibard, \href{https://www.nature.com/articles/nature04851}{Nature {\bf 441} 1118 (2006)}.

\bibitem{Hung} Ch-L. Hung, Xibo Zhang, Nathan Gemelke and Cheng Chin, \href{https://www.nature.com/articles/nature09722} {Nature {\bf 470}, 236 (2011).}

\bibitem{U2D} A. Posazhennikov, \href{https://journals.aps.org/rmp/abstract/10.1103/RevModPhys.78.1111}{Rev. Mod. Phys. \textbf{78}, 1111 (2006)}. 

\bibitem{Salasnich} L. Salasnich, A. Parola, L. and Reatto, \href{https://journals.aps.org/pra/abstract/10.1103/PhysRevA.65.043614}{Phys. Rev. A {\bf 65}, 043614 (2002)}.

\bibitem{Mateo} A. M. Mateo and V. Delgado, \href{https://journals.aps.org/pra/abstract/10.1103/PhysRevA.77.013617}{Phys. Rev. A {\bf 77}, 013617 (2008)}.

\bibitem{Bao} W. Bao, D. Jaksch, and P. A. Markowich, \href{https://www.sciencedirect.com/science/article/pii/S0021999103001025}{J. of Comput. Phys. {\bf 187} 318 (2003)}.

\bibitem {Trallero} C. Trallero-Giner, R. Cipolatti, and T. C. H. Liew, \href{https://link.springer.com/article/10.1140/epjd/e2013-40163-9}{ Eur. Phys. J. D. {\bf 67} 143 (2013)}.

\bibitem{Zamora} R. Zamora-Zamora R, G. A. Dom\'{\i}nguez-Castro, C. Trallero-Giner, R. Paredes and V. Romero-Roch\'{\i}n, \href{https://iopscience.iop.org/article/10.1088/2399-6528/ab360f}{J. Phys. Comm. {\bf 3} 085003 (2019)}.

\bibitem{U2D2}
D. S. Petrov, M. Holzmann and G. V Shlyapnikov, \href{https://journals.aps.org/prl/abstract/10.1103/PhysRevLett.84.2551}{Phys. Rev. Lett. \textbf{84}, 2551 (2000)}.

\bibitem{Lung-Hung} C.-L. Hung , X. Zhang, N. Gemelke, and C. Chin, \href{https://www.nature.com/articles/nature09722}{Nature {\bf 470} 236 (2011)}.

\bibitem{Wang} F. Wang, X. Li, D. Xiong, and D. Wang, \href{https://iopscience.iop.org/article/10.1088/0953-4075/49/1/015302/meta}{J. Phys. B: At. Mol. Opt. Phys. {\bf 49} 015302 (2016)}.

\bibitem{Papp} S. B. Papp, J. M. Pino, and C. E. Wieman, \href{https://journals.aps.org/prl/abstract/10.1103/PhysRevLett.101.040402}{Phys. Rev. Lett. {\bf 101} 040402 (2008)}.

\bibitem{Ray} S. Ray, M. Pandey, A. Ghosh, and S. Sinha, \href{https://iopscience.iop.org/article/10.1088/1367-2630/18/1/013013}{New. J. Phys. \textbf{18}, 013013 (2016)}. 

\bibitem{Schulte} T. Schulte, S. Drenkelforth, J. Kruse, R. Tiemeyer, K. Sacha, J. Zakrzewski, M. Lewenstein, W. Ertmer, and J. J. Arlt, \href{https://iopscience.iop.org/article/10.1088/1367-2630/8/10/230}{New. J. Phys. \textbf{8}, 230 (2006)}. 

\bibitem{Adhikari} S. K. Adhikari and L. Salasnich, \href{https://journals.aps.org/pra/abstract/10.1103/PhysRevA.80.023606}{Phys. Rev. A \textbf{80}, 023606 (2009)}. 

\bibitem {Michelangeli} A. Michelangeli and A. Olgiati, \href{https://www.atlantis-press.com/journals/jnmp/125950572} { J. of Nonlinear Mathematical Physics {\bf 24},3 426 (2017)}.

\bibitem{CMadronero} C. Madro\~nero, G.A. Dom\'{\i}nguez-Castro, L. A. Gonz\'alez-Garc\'{\i}a and R. Paredes. \href{https://journals.aps.org/pra/abstract/10.1103/PhysRevA.102.033304}
{Phys. Rev. A, {\bf 102} 033304, (2020)}.

\bibitem{Calculations} The grid used in our numerical calculations was 512 $\times$ 512. All the calculations were implemented in state of the art GPU hardware.

\bibitem{Dum} R. Dum, Y. Castin, \href{https://link.springer.com/article/10.1007/s100530050584}{Eur. Phys. J. D {\bf 7}, 399 (1999)}.

\bibitem{Zhang} R. Zeng, Y. Zhang, \href{https://www.sciencedirect.com/science/article/pii/S0010465508004207}{Comput. Phys. Commun. {\bf 180}, 854 (2008)}.

\bibitem{Weld}  D. M. Weld, P. Medley, H. Miyake, D. Hucul, D. E. Pritchard, and W. Ketterle, \href{https://journals.aps.org/prl/abstract/10.1103/PhysRevLett.103.245301} {Phys. Rev. Lett. {\bf 103}, 245301 (2009)}.

\bibitem{Feuerbacher} M. Feuerbacher, \href{https://scripts.iucr.org/cgi-bin/paper?S2053273321007245} {Acta Cryst. {\bf A77}, 460 (2021)}.

\bibitem{Palacios} S. Palacios Alvarez, P. Gomez, S. Coop. R. Zamora-Zamora. C. Mazzinghi and M. Mitchel, \href{https://www.pnas.org/doi/10.1073/pnas.2115339119} {PNAS {\bf 119} 6, 5339 (2022)}.

\bibitem{Kennes} D. M. Kennes, M. Claassen, L. Xian, A. Georges, A. J. Millis, J. Hone, C. R. Dean, D. N. Basov, A. N. Pasupathy and A. Rubio \href{https://www.nature.com/articles/s41567-020-01154-3}{Nat. Phys., {\bf 17}, 155 (2021)}.

\end{thebibliography}
\end{document}